\documentclass[prl,aps,11pt]{revtex4}
\usepackage{amsmath}
\usepackage{amssymb}
\usepackage{epsfig}
\usepackage{amscd}

\begin{document}

\title{Hadamard NMR spectroscopy for two-dimensional quantum information processing and parallel search algorithms}
\author{T. Gopinath and Anil Kumar\footnote{Electronic address: anilnmr@physics.iisc.ernet.in}}
\affiliation{{\small \it NMR Quantum Computing and Quantum Information Group.\\Department of Physics and NMR Research Centre.\\
Indian Institute of Science, Bangalore - 560012, India.}}

\begin{abstract}
Hadamard spectroscopy has earlier been used to speed-up multi-dimensional NMR experiments. 
In this work we speed-up the two-dimensional quantum computing scheme, by using Hadamard spectroscopy in the indirect 
dimension, resulting in a scheme which is faster and 
requires the Fourier transformation only in the direct dimension. 
Two and three qubit quantum gates are implemented with an extra observer qubit. 
We also use one-dimensional Hadamard 
spectroscopy for binary information storage by spatial encoding and implementation of a parallel search algorithm. 
\end{abstract}

\maketitle

\section{I. Introduction}
The use of quantum systems for information processing was first introduced by Benioff \cite{beni}. In 1985 Deutsch described quantum computers 
which exploit the superposition of multi particle states, thereby achieving massive parallelism \cite{deu}.  
Researchers have also studied the possibility of solving certain types of problems more efficiently than can be done on conventional computers 
\cite{deujoz,pw,gr}. 
These theoretical possibilities have generated significant interest for experimental realization of quantum computers \cite{ic,benson}. 
Several techniques are being exploited for quantum computing and quantum information processing including nuclear magnetic resonance (NMR) \cite{cor,ger}. 
 
NMR has played a leading role for the practical demonstration of quantum gates and algorithms \cite{cor1,nmr1,nmr2}. 
In NMR, individual spins having different Larmor frequencies and weakly coupled to each other are treated as individual 
qubits. 
The 
unitary operators needed 
for the implementation, have mostly been realized using spin selective as well as transition selective radio frequency pulses and coupling evolution, 
utilizing spin-spin (J) or dipolar couplings among the spins \cite{nmr3,nmr4,kavita,mahesh}. The final step of any quantum 
computation is the read out of the output. In NMR the read 
out is obtained by selectively detecting the magnetization of each spin or by tomography of full density matrix 
\cite{cor2,rana1}. 
It was first proposed by Ernst and co-workers, that a two dimensional experiment can be used to correlate, input and output states, 
which is advantageous 
from spectroscopic view point \cite{madi}. In two-dimensional quantum computation (2D QC), an extra qubit (observer qubit) is used, 
whose spectral lines indicate the quantum states of the work qubits \cite{madi}. The 2D spectrum of the observer qubit, gives 
input-output correlation of the computation performed on the work qubits \cite{madi}. 
The 2D 
spectrum is therefore more informative than a one dimensional (1D) 
spectrum. For example, in 1D NMR QIP, the spectrum after the SWAP operation, performed on the equilibrium state of a homonuclear system, is identical to 
the equilibrium spectrum. However the same operation, performed using a 2D experiment, contains the signature of SWAP gate \cite{nmr5}. The observer qubit can also be used to prepare a pair of pseudo pure states \cite{rana2}. The quantum logic gates 
and several algorithms are implemented by 2D NMR \cite{mahesh,nmr5,rana2}. Recently, 2D NMR has also been used to address the decoherence free sub spaces, for 
quantum information processing \cite{wei}.

Multi-dimensional NMR spectroscopy is often time consuming, since each indirect dimension has to be incremented to span the whole frequency 
range, and 
the desired digital resolution \cite{ern}. 
Several experimental protocols have been developed to accelerate the recording of multi-dimensional 
spectra. These 
include, single scan experiments in the presence of large gradients, GFT, Covariance spectroscopy and Hadamard spectroscopy 
\cite{frydman,thomas,brusch,kupce2,kupce3,kupce4,kupce5}. The 
Hadamard spectroscopy, proposed by Kupce and Freeman, has the advantage that 
one can simultaneously label, various transitions of the spectrum by applying a multi-frequency pulse \cite{kupce2,kupce3}. A suitable decoding followed 
by a Fourier 
transform only in the direct dimension yields a 2D spectrum \cite{kupce2,kupce3}. This leads to a large saving in time for experiments having a 
small number of 
transitions \cite{kupce2,kupce3}. In this paper we use Hadamard spectroscopy to speed-up the two dimensional quantum computing scheme \cite{madi}.

Information storage and retrieval at the atomic and molecular level has been an active area of research 
\cite{anderson,eigler,eigler1,sersa1,sersa2,sersa3,kiruluta,khitrin}. 
Khitrin et.al demonstrated that, the $^1{H}$ 
spectrum of dipolar coupled spin cluster can be used to store large amounts of information, which can be used for 
photography and implementation of parallel search algorithm \cite{khitrin,khitrin1,fung,khitrin2}. Alternately, 
it has been demonstrated that, spatial encoding under a linear field gradient can also be used for above purposes 
\cite{dey,rangeet}. 
In this work, the one-dimensional Hadamard spectroscopy \cite{kupce1} is used under spatial encoding, to store the information and to implement a 
parallel search algorithm. 
The proposed method has the advantage that, once the Hadamard encoded data is recorded, one can write any binary information array (sentence), and search 
for any 
code or alphabet in that array. The main emphasis of this paper is to demonstrate the use of Hadamard encoding in the 
field of NMR information processing.

In section (II), we outline the Hadamard method for 2D-NMR QIP along with the conventional method. In section (III) we implement, various 2D-gates 
on 3 and 4-qubit systems. In section (IV), we implement 
parallel search algorithm by using Hadamard spectroscopy under spatial encoding and section (V) contains the conclusions.

\section{II. Theory}

Quantum computing using two-dimensional NMR can be described by transformations in the Liouville space. For a spin-1/2 nucleus, having two orthogonal 
states $\vert 0 \rangle$ and $\vert 1 \rangle$, the longitudinal polarization operators can be written as \cite{ern,madi},

\begin{eqnarray}
I_{0}= \vert 0 \rangle \langle 0 \vert = \begin{pmatrix} 1&0\cr 0&0 \end{pmatrix},\hspace{0.2cm} 
I_{1}= \vert 1 \rangle \langle 1 \vert = \begin{pmatrix} 0&0\cr 0&1 \end{pmatrix},\hspace{0.2cm} 
and \hspace{0.2cm} I_z = \frac {1}{2}(I_{0}-I_{1})= \frac {1}{2} \begin{pmatrix} 1&0\cr 0&-1 \end{pmatrix}
\end{eqnarray}

A product state $\vert \psi \rangle$= $\vert 001...0 \rangle$ of a coupled spin-1/2 nuclei, can be represented in the Liouville space by a density 
matrix $\sigma$, obtained by the 
direct product of longitudinal operators, 
\begin{eqnarray}
\sigma= I_0 \otimes I_0 \otimes I_1 \otimes ....\otimes I_0  =I_0I_0I_1....I_0.
\end{eqnarray}

 In 2D-NMR QIP \cite{madi}, an extra qubit (observer qubit) is used, whose transitions represent the quantum states of the work qubits (computation qubits). Thus an 
(N+1)-qubit system can be used for N-qubit computation, treating the zeroth qubit as the observer qubit. 
The thermal equilibrium state of observer spin $I^O$, can be represented in the Liouville space as \cite{madi},
\begin{eqnarray}
\sigma_{eq}^O= I_z^O[(I_0^1I_0^2....I_0^N)+(I_0^1I_0^2....I_1^N)+.........+(I_1^1I_1^2....I_1^N)],
\end{eqnarray}
where the superscript indicates the qubit number, with the observer qubit represented by the letter 'O'.

In the following we describe the conventional and the Hadamard 2D methods (Fig.\ref{pulse sequence}), for a three qubit system,
under the NOT operation on both the work qubits, during the computation period.
The schematic energy level diagram of a three qubit system and the spectrum of the observer 
qubit are given in Fig.\ref{energy}. The transitions of the observer qubit, which represent the quantum states of the other two qubits (work qubits), 
are labeled as 
$\vert 00 \rangle$, $\vert 01 \rangle$, $\vert 10 \rangle$ and $\vert 11 \rangle$. A NOT operation performed during the computation period, interchanges 
the states $\vert 0 \rangle$ and $\vert 1 \rangle$ of both the work qubits.

\subsection{(A) The Conventional Method}

As shown in Fig.(\ref{pulse sequence}A), the observer spin is first allowed to evolve for a time $t_1$ during which the work 
qubits remain in their initial states, after the frequency labeling period $t_1$, the computation is performed on the work qubits, followed by the
detection in $t_2$ period. A two-dimensional Fourier transform gives the 2D spectrum of the observer qubit, which shows the input-output correlation of 
the computation, performed on the work qubits. 

For a three qubit system, the equilibrium state of the observer qubit $I^O$, can be written as,
\begin{eqnarray}
\sigma_z^{eq}= I_z^O[(I_0^1I_0^2)+(I_0^1I_1^2)+(I_1^1I_0^2)+(I_1^1I_1^2)].
\end{eqnarray}

The pulse sequence given in fig.\ref{pulse sequence}A, transforms $\sigma_z^O$ as,

\begin{equation}
\begin{split}
\sigma_z^O \xrightarrow[\hspace*{1cm}]{(\pi/2)_y}\;\; &  I_x^O[(I_0^1I_0^2)+(I_0^1I_1^2)+(I_1^1I_0^2)+(I_1^1I_1^2)]\cr
\xrightarrow[\hspace*{1cm}]{t_1} \;\;&  I_x^O[cos(\omega_{00}t_1)(I_0^1I_0^2)+cos(\omega_{01}t_1)(I_0^1I_1^2)+cos(\omega_{10}t_1)(I_1^1I_0^2)+cos(\omega_{11}t_1)(I_1^1I_1^2)]\cr
\xrightarrow[\hspace*{1cm}]{(\pi/2)_{-y}^{I^O}} \;\;& I_z^O[cos(\omega_{00}t_1)(I_0^1I_0^2)+cos(\omega_{01}t_1)(I_0^1I_1^2)+cos(\omega_{10}t_1)(I_1^1I_0^2)+cos(\omega_{11}t_1)(I_1^1I_1^2)]\cr
\xrightarrow[\hspace*{1cm}]{U_{NOT}} \;\;& I_z^O[cos(\omega_{00}t_1)(I_1^1I_1^2)+cos(\omega_{01}t_1)(I_1^1I_0^2)+cos(\omega_{10}t_1)(I_0^1I_1^2)+cos(\omega_{11}t_1)(I_0^1I_0^2)]\cr
\xrightarrow[\hspace*{1cm}]{(\pi/2)_y} \;\;& I_x^O[cos(\omega_{00}t_1)(I_1^1I_1^2)+cos(\omega_{01}t_1)(I_1^1I_0^2)+cos(\omega_{10}t_1)(I_0^1I_1^2)+cos(\omega_{11}t_1)(I_0^1I_0^2)]\cr
\xrightarrow[\hspace*{1cm}]{t_2} \;\;& I_x^O[cos(\omega_{00}t_1)cos(\omega_{11}t_2)(I_1^1I_1^2)+cos(\omega_{01}t_1)cos(\omega_{10}t_2)(I_1^1I_0^2) \cr 
 & + cos(\omega_{10}t_1)cos(\omega_{01}t_2)(I_0^1I_1^2)+cos(\omega_{11}t_1)cos(\omega_{00}t_2)(I_0^1I_0^2)] 
\end{split}
\end{equation}

Fourier transform performed in both dimensions on the above signal, gives a two dimensional spectrum, where input and output states are given in $F_1$ and $F_2$ dimensions 
respectively. The time consuming part of this method is the large number of $t_1$ increments, needed to achieve the required spectral width and 
sufficient resolution in the $F_1$ dimension. Quadrature detection in the $F_1$ dimension, further doubles the number of experiments. 

\subsection{(B) The Hadamard Method}

In this method (Fig.\ref{pulse sequence}B), the sequence $(\pi/2)$-$t_1$-$(\pi/2)-G_z$ of Fig.\ref{pulse sequence}A, is replaced by a 
multi-frequency $\pi$ (MF-$\pi$) pulse on the observer qubit. 
Instead of $t_1$ increments of method (A), the pulse sequence of Fig.\ref{pulse sequence}B, is repeated k times, where 
$k=2^N$, is the number of transitions of the observer 
qubit. In each of the k experiments, the multi-frequency $\pi$ pulse is differently encoded, according to the rows of a k-dimensional Hadamard 
matrix. 
For a two work qubit case (k=4), four experiments are performed with four different encodings of the $\pi$ pulse, given by the four rows of the four 
dimensional 
Hadamard matrix (Fig. \ref{hada4}a), where '-' and '+' in the matrix corresponds to '$\pi$ pulse' and 'no $\pi$ pulse' respectively. For example, + - + - means, 
the  $\pi$ 
pulse 
is applied only on 2nd and 4th transitions of the observer qubit. The output of the four experiments (Fig.\ref{hada4}a), under the NOT operation on both the work qubits, can be 
calculated as follows,

Experiment (1):
\begin{eqnarray}
\begin{CD}
\sigma_z^o @>no pulse>> I_z^O[(I_0^1I_0^2)+(I_0^1I_1^2)+(I_1^1I_0^2)+(I_1^1I_1^2)]\\
@>U_{Not}>> I_z^O[(I_1^1I_1^2)+(I_1^1I_0^2)+(I_0^1I_1^2)+(I_0^1I_0^2)]\\
@>(\pi/2)-t>> I_x^O[cos(\omega_{11}t)(I_1^1I_1^2)+cos(\omega_{10}t)(I_1^1I_0^2)+cos(\omega_{01}t)(I_0^1I_1^2)+cos(\omega_{00}t)(I_0^1I_0^2)];
\end{CD}
\end{eqnarray}

Experiment (2):
\begin{eqnarray}
\begin{CD}
\sigma_z^O @>(\pi)^{\vert01\rangle, \vert11\rangle}>> I_z^O[(I_0^1I_0^2)-(I_0^1I_1^2)+(I_1^1I_0^2)-(I_1^1I_1^2)]\\
@>U_{Not}>> I_z^O[(I_1^1I_1^2)-(I_1^1I_0^2)+(I_0^1I_1^2)-(I_0^1I_0^2)]\\
@>(\pi/2)-t>> I_x^O[cos(\omega_{11}t)(I_1^1I_1^2)-cos(\omega_{10}t)(I_1^1I_0^2)+cos(\omega_{01}t)(I_0^1I_1^2)-cos(\omega_{00}t)(I_0^1I_0^2)];
\end{CD}
\end{eqnarray}

Experiment (3):
\begin{eqnarray}
\begin{CD}
\sigma_z^o @>(\pi)^{\vert10\rangle, \vert11\rangle}>> I_z^O[(I_0^1I_0^2)+(I_0^1I_1^2)-(I_1^1I_0^2)-(I_1^1I_1^2)]\\
@>U_{Not}>> I_z^O[(I_1^1I_1^2)+(I_1^1I_0^2)-(I_0^1I_1^2)-(I_0^1I_0^2)]\\
@>(\pi/2)-t>> I_x^O[cos(\omega_{11}t)(I_1^1I_1^2)+cos(\omega_{10}t)(I_1^1I_0^2)-cos(\omega_{01}t)(I_0^1I_1^2)-cos(\omega_{00}t)(I_0^1I_0^2)];
\end{CD}
\end{eqnarray}

\begin{center}
\begin{figure}
\epsfig{file=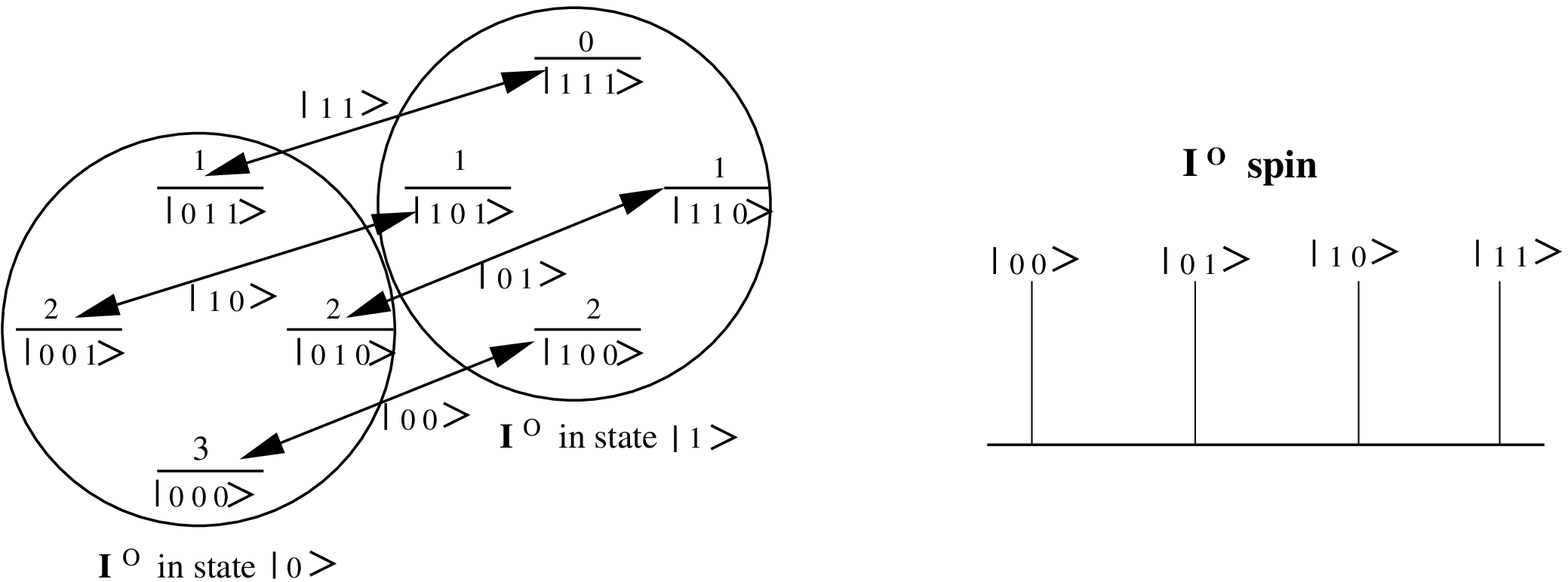,width=15cm}
\caption{} \label{energy} 
\end{figure}
\end{center}

Experiment (4):
\begin{eqnarray}
\begin{CD}
\sigma_z^o @>(\pi)^{\vert01\rangle, \vert10\rangle}>> I_z^O[(I_0^1I_0^2)-(I_0^1I_1^2)-(I_1^1I_0^2)+(I_1^1I_1^2)]\\
@>U_{Not}>> I_z^O[(I_1^1I_1^2)-(I_1^1I_0^2)-(I_0^1I_1^2)+(I_0^1I_0^2)]\\
@>(\pi/2)-t>> I_x^O[cos(\omega_{11}t)(I_1^1I_1^2)-cos(\omega_{10}t)(I_1^1I_0^2)-cos(\omega_{01}t)(I_0^1I_1^2)+cos(\omega_{00}t)(I_0^1I_0^2)];
\end{CD}
\end{eqnarray}

where $(\pi)^{\vert ij \rangle, \vert lm \rangle}$ means, a $\pi$ pulse is applied on $\vert ij \rangle$ and 
$\vert lm \rangle$ transitions of the observer qubit.
Each of the four experiments (eqn.s 6 to 9) generates a composite response of the computation, performed on the work qubits. However, the different 
encoding pattern 
applied in 
each experiment, provides a decoding method of extracting the output state, individually for each of the input states.
The decoding is obtained by the transpose of the Hadamard matrix. 
The decoding of the experiments 1, 2, 3 and 4, for the input states  (Fig.\ref{energy}b), $\vert 00 \rangle$, $\vert 01 \rangle$, $\vert 10 \rangle$ and 
$\vert 11 \rangle$ are respectively given by,

\begin{eqnarray}
(1)+(2)+(3)+(4),  \nonumber  \\ 
(1)-(2)+(3)-(4), \\ \nonumber 
(1)+(2)-(3)-(4), \\ \nonumber 
 (1)-(2)-(3)+(4). 
\end{eqnarray}

A two-dimensional spectrum of the computation can be constructed, by inserting the decoded data (time domain) at suitable frequencies in the $F_1$ 
dimension following a Fourier transform in the $F_2$ dimension \cite{kupce2}. 
For example, in the present case, the decoded data (eqn.10) for the input states $\vert 00 \rangle$, $\vert 01 \rangle$, $\vert 10 \rangle$ and $\vert 11 \rangle$
 are inserted in the $F_1$ dimension, respectively at the frequencies 
$\omega_{00}$, $\omega_{01}$, $\omega_{10}$ and $\omega_{11}$ (Fig. \ref{energy}), following a Fourier transform in the $F_2$ dimension, yielding the desired 2D spectrum.

 It is to be noted that, the Hadamard encoding can also be achieved by J-evolution. For example in the above case (eqns. 6 to 9), the observer qubit can be represented in terms of product operators (respectively for eqn.s 6 to 9, fig.\ref{hada4}a) as, 
$I_z^O$, ($I_z^OI_z^2$), ($I_z^OI_z^1$), and ($I_z^OI_z^1I_z^2$), after the Hadamard encoding (MF-$\pi$ pulse). Each of these product operators can also be prepared by using J-evolution method \cite{nmr3}.

While the conventional method (Fig.\ref{pulse sequence}A) needs a minimum number of $t_1$ increments for a satisfactory 
resolution in the $F_1$ dimension, the Hadamard method (Fig.\ref{pulse sequence}B) inherently yields high resolution in $F_1$ 
dimension and needs only a small number of experiments, equal to number of 
transitions of the observer qubit. It may be noted that for an N work qubits system, the number of 
transitions of the observer qubit (for weakly coupled spins with all resolved transitions) is $2^N$, thus for small number 
of qubits (up to about 9 qubits) the Hadamard method is advantageous. 
The Hadamard method can also be used for 2D implementation of 
quantum algorithms \cite{nmr5,rana2}. It may be added that the Hadamard method does not change the scaling of quantum 
computing nor does it change the scaling of any algorithm.

\section{III. Experimental implementation of quantum gates}

\subsection{(a) Two qubit gates} 

The system chosen for implementation of two qubit gates, is $C_2F_3I$, where the three fluorines can be treated as three qubits. The fluorine spectra are shown in 
Fig. \ref{1D of 3qubits}. The 
transitions of the observer qubit $I^O$ are labeled as $\vert 00 \rangle$, $\vert 01 \rangle$, $\vert 10 \rangle$ and $\vert 11 \rangle$.

 The two-qubit NOT(1,2) gate (NOT on qubits 1 and 2) is implemented (Fig. \ref{2D of 3qubits}a) using method (A), with 
128 $t_1$ increments, a recycle delay of 20 seconds ($\sim$$5T_1$) 
and 2 scans for each increment, resulting in a 
total experimental time of 126 minutes.
 The two-qubit NOT(1,2) and several other gates are implemented by method (B), shown in Fig.\ref{2D of 3qubits}b. The 
Hadamard encoding, shown in Fig.\ref{hada4}a, is achieved 
by MF-$\pi$  pulses of duration 
100 ms.. 
The unitary operators and 
computation pulses for various two-qubit gates are given in \cite{nmr5}. 
The NOP gate is a unit matrix, hence the output states are same as input states,
NOT(1,2) interchanges $\vert 0 \rangle$ and $\vert 1 \rangle$ of both the work qubits, 
SWAP gate interchanges the states $\vert  01 \rangle$ and $\vert  10 \rangle$, and CNOT(1) gate interchanges the states
$\vert  01 \rangle$ and $\vert  11 \rangle$.
Each 2D gate of Fig.\ref{2D of 3qubits}b, is recorded in four experiments, 
which for same recycle delay as in method (A), takes the total experimental time of less than 2 minutes.

\subsection{(b) Three qubit gates}

The four fluorines of 2-amino, 3,4,5,6-tetra fluoro benzoic acid, can be used as four qubits.
The one-dimensional spectra of observer qubit $(I^O)$ and  work qubits ($I^1$, $I^2$ and $I^3$) are given in (Fig.\ref{1D of 4qubits}),

Since for this system the $(I^O)$ 
spin has 8 transitions the Hadamard method (Fig.\ref{pulse sequence}B) for 2D QC gates requires 8 experiments as outlined 
in Fig.\ref{hada4}b. 
Due to small separation of the frequencies, Fig.\ref{1D of 4qubits} 
( 5 Hz, as compared to 40 Hz 
in Fig. \ref{1D of 3qubits}), 
the MF-$\pi$ pulse needs about 600 ms.. Hence the Hadamard encoding, in this case, is achieved by using J-evolution 
method \cite{nmr3}, explained below.

The magnetization of the observer qubit, after the Hadamard encoding (Fig.\ref{pulse sequence}b), can be represented as, 
$I_z^O$, $I_z^OI_z^3$, $I_z^OI_z^1$, $I_z^OI_z^1I_z^3$, $I_z^OI_z^2$, $I_z^OI_z^2I_z^3$, $I_z^OI_z^1I_z^2$, $I_z^OI_z^1I_z^2I_z^3$ 
(Fig.\ref{hada4}b). The pulse sequence
$(\pi/2)_y^O$-$(1/2J_{Oi})$-$(\pi/2)_x^O$ is used to prepare $I_z^OI_z^i$, where the evolution is with respect to 
$J_{Oi}$. The product operator $I_z^OI_z^iI_z^j$ $(i \ne j)$ is prepared by the pulse sequence
$(\pi/2)_y^O$-$(1/2J_{Oi})$-$(1/2J_{Oj})$-$(\pi/2)_y^O$, and $I_z^OI_z^1I_z^2I_z^3$ is prepared by the pulse sequence,
$(\pi/2)_y^O$-$(1/2J_{O1})$-$(1/2J_{O2})$-$(1/2J_{O3})$-$(\pi/2)_{-x}^O$. Each $1/2J_{Oi}$ evolution is achieved by applying 
selective $\pi$ pulses simultaneously on the observer and the $i^{th}$ qubit, in the middle of the evolution period 
$(1/2J_{Oi})$.

A NOP gate is implemented, which requires no operation during the computation, is shown in Fig.\ref{2D of 4qubits}a. NOT(1) and NOT(2) are implemented
by applying a $\pi$ pulse, respectively on $I^1$ and $I^2$ (Fig.\ref{2D of 4qubits}b,c). The Toffoli gate, which is a universal gate for reversible
computation, is achieved by applying two transition selective $\pi$ pulses on transitions 
$\vert 0110 \rangle$-$\vert 0111 \rangle$ and $\vert 1110 \rangle$-$\vert 1111 \rangle$ (\ref{2D of 4qubits}d). 
Each 2D gate shown in Fig.\ref{2D of 4qubits}, takes about 2 minutes, for a recycle delay of 8 seconds. The conventional 
method with 128 $t_1$ increments, takes about 60 minutes (spectrum not shown).

\section{IV. Parallel search algorithm}

Information storage by NMR, was suggested almost 50 years ago by Anderson et.al \cite{anderson}. This involves the excitation of various slices of the isotropic liquid (for example $H_2O$), 
under the z-gradient, using series of weak radio-frequency pulses followed by spin echo \cite{anderson}. 
Khitrin et.al demonstrated that, the multi-frequency excitation of dipolar coupled 
$^{1}H$ spectrum of liquid crystal, enables a parallel (simultaneous) storage of the information, at the atomic level 
\cite{khitrin}. In ref.\cite{khitrin,khitrin1}, it is shown that, one can imprint 
the information 
written in a binary code, where '0' and '1', in the frequency space corresponds to no excitation and excitation respectively. A 215 bit sentence is 
written, where each alphabet is assigned a five bit string, for example a=1 (00001),........z=26(11100) and blank space as 00000 \cite{khitrin}. 
It is further shown that, one can perform a parallel search on the binary information array 
(sentence) using six bit-shifted multi-frequency pulses, 
to search for a letter, in a string of letters \cite{khitrin2}.

 Recently, it has been demonstrated that, spatial encoding can also be used for information storage and parallel search using the single resonance of a 
liquid such as $H_2O$ in the presence of a linear field gradient \cite{dey,rangeet}. Spatial encoding involves 
radio-frequency (rf) excitation at multiple frequencies in the presence of a linear magnetic 
field gradient along 
the z-direction. Spatial encoding was also used by Sersa et.al, to excite an arbitrary three 
dimensional patterns, using x, y, and z gradients \cite{sersa1,sersa2,sersa3}. 
NMR photography and a parallel 
search algorithm using XOR operation have been implemented by spatial encoding under z-gradient \cite{dey,rangeet}. 
The XOR search requires only two experiments, in which 
the first experiment is used to record the sentence and the second to record the ancilla pattern of the letter to be searched. 
The absolute intensity 
difference spectrum of these two experiments is obtained, followed by 
the integration of the peak intensities of the 5 bits corresponding to each of the letters. 
The pattern of the integrated intensities yields the zero intensity, only at those places where the required letter is present. The XOR search not only 
searches 
the required letter, but also searches the letter having the complementary code. For example, as shown in \cite{rangeet}, the letters "o = 01111" and it's 
complementary "p=10000" 
can be searched simultaneously. 

 In this work, the spatial encoding is conjugated with Hadamard spectroscopy \cite{kupce1}. 
We synthesize 256 phase encoded multi-frequency $\pi/2$ pulses, each of which consists of 256 harmonics and excite the 256 slices of the water sample, 
under z-gradient (fig.\ref{pulse sequence for spatial encoding}). The phase encoding of the harmonics, in each 
of the pulses, 
is given by the rows of the 256 dimensional Hadamard matrix, where '+' and '-' in the matrix, corresponds to the phases y and -y respectively. 
Thus the application of the pulse under the gradient (fig.\ref{pulse sequence for spatial encoding}), creates either $I_x$ or -$I_x$ magnetization of each slice. 
The 256 pulses are used to record the Hadamard encoded data of 256 slices, by using the pulse sequence of fig.\ref{pulse sequence for spatial encoding}.  
As seen in the previous sections, the Hadamard encoded data can be suitably decoded to obtain any element (frequency) of the Hadamard matrix. 
It should be noted that, in this case the Hadamard encoding is performed on the $I_x$ magnetization of various slices, whereas in the 2D gates 
(Fig.\ref{pulse sequence}b, section (II)), the encoding is done on the z magnetization of the observer qubit transitions.
The Hadamard encoded data of 256 slices, stored in 256 separate files, can be suitably decoded to write any binary information, which requires a maximum of 256 bits.

 The Hadamard encoded data is suitably decoded, to write the sentence, "the quick brown fox jumps over the lazy dog" (Fig.\ref{fox}A,B). The XOR search 
\cite{rangeet} is 
performed to 
search a letter "u", the ancilla pattern for the letter "u" is 
decoded in Fig.(\ref{fox}C), and the difference spectrum of Figs.\ref{fox}A and \ref{fox}C, is shown in Fig(\ref{fox}D). 
Integration of absolute intensities of the peaks 
of each of the letters in Fig.(\ref{fox}D), are 
shown in Fig.(\ref{fox}E). The zero intensity in Fig.(\ref{fox}E), indicates the presence of letter "u". 
Maximum intensity is observed for the letter "j", which has a code complementary to letter "u". 
The advantage of method is seen in the next example, in which the same data is used to write another sentence and search 
code for another letter. Only the decoding pattern is different in example of 
Figs.\ref{pnmr}A,
contains another sentence, "principles of nuclear magnetic resonance", which consists of 200 bits of information. 
The letter "e" 
is searched by using the XOR search. The encoded data can also be used for 256 bits of NMR photography \cite{khitrin1,fung,dey,kiruluta}.
The spatial encoding \cite{dey,rangeet} has the advantage that the relaxation of all the slices (bits) is uniform. It may be pointed out that in the liquid crystal method \cite{khitrin2} as well as
J-coupled systems \cite{fung1} all the
lines are not independent and perturbation of one line can cause disturbance in other lines of the spectrum, which
forms the basis of the Z-COSY experiment \cite{grace}. On the
other hand, in spatial encoding method the pattern is inhomogeneous broadened and parts of the spectrum can be
independently perturbed \cite{dey,rangeet}.

\section{V. Conclusions}

In this paper we demonstrate the use of Hadamard encoding for (i) two-dimensional quantum information processing and (ii) for parallel search using 
spatial encoding. 
For (i), this method converts the 2D experiment to 
a small number of 1D experiments requiring the Fourier 
transformation, only in the direct dimension. The required encoding can be achieved by using multi-frequency $\pi$  pulses or J-evolution method. 
For (ii) the Multi-frequency excitation and detection of the water sample, in the presence of z-gradient, maps the 
magnetization of various slices, to the frequency space. Each slice is treated as a classical bit which can exists either in 0 or 1, which in the frequency 
space respectively correspond to no excitation and excitation. The Hadamard encoded data is suitably decoded for the 
information storage and implementation of parallel search algorithms. It will be interesting to use the Hadamard 
spectroscopy, for information storage using x,y and z gradients.

\section{ACKNOWLEDGMENTS}

Useful discussions with Dr. Rangeet Bhattacharyya and Raghav G Mavinkurve are gratefully acknowledged. The use of AV-500 NMR spectrometer funded by the 
Department of
Science
and Technology (DST), New Delhi, at the NMR Research Centre, Indian Institute of Science,
Bangalore, is gratefully acknowledged. A.K. acknowledges DAE for Raja Ramanna Fellowship, and DST for a research grant on
"Quantum Computing using NMR techniques".

\pagebreak

\pagebreak
\section{Figure captions}

\begin{center}
\begin{figure}
\epsfig{file=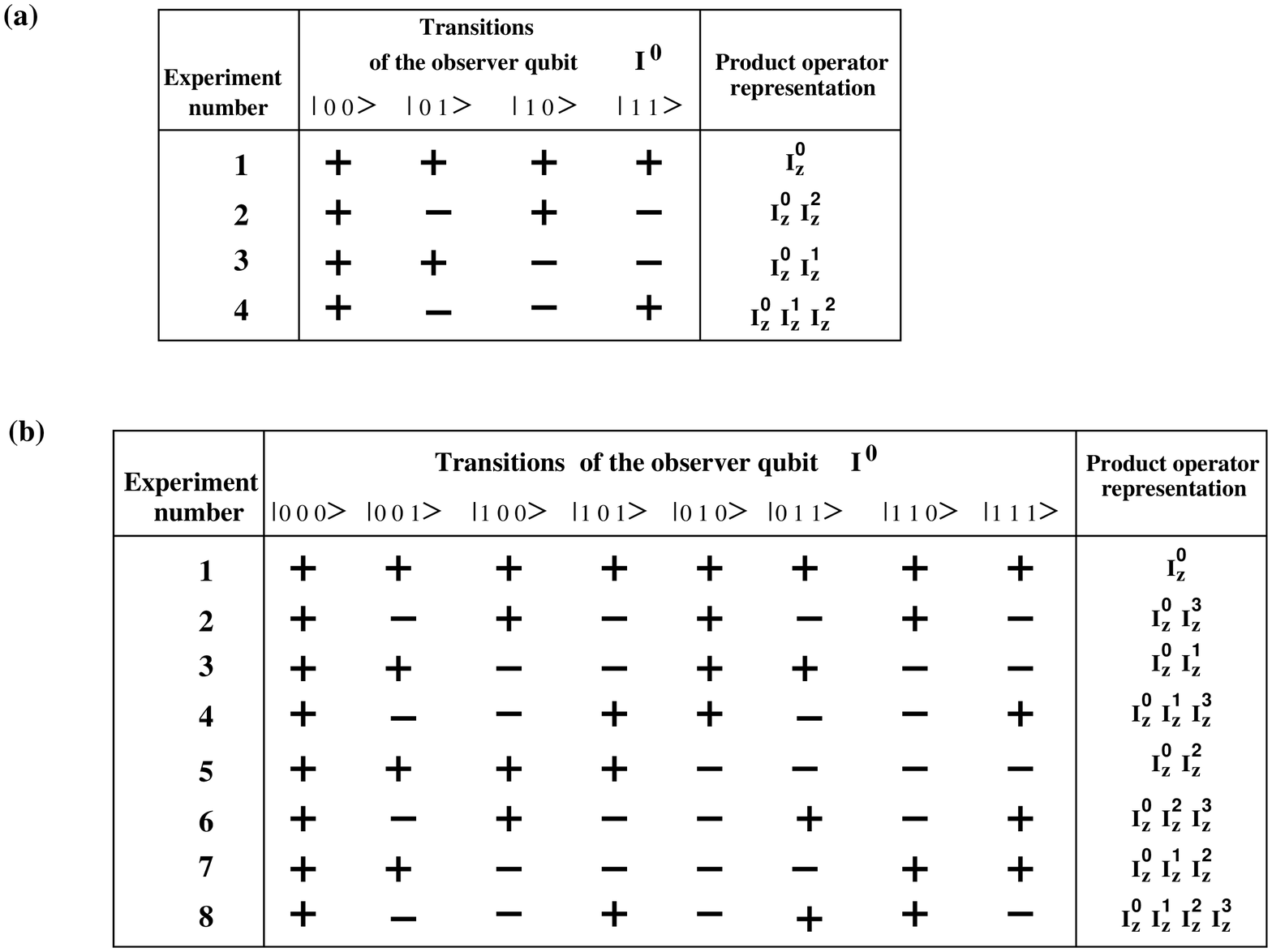,width=18cm}
\caption{}   \label{hada4}
\end{figure}
\end{center}

(1).(a) Pulse sequence for 2D NMR QIP. $I^O$ is the observer qubit, and $I^1$, $I^2$,....$I^N$ are work qubits. During $t_1$ period the input states of the 
work qubits are labeled followed by the computation, and signal acquisition during the $t_2$ period. A two dimensional Fourier transform results the 2D spectrum 
of the observer qubit, where the input and output states are given in $F_1$ and $F_2$ dimensions respectively.
(b) pulse sequence for 2D Hadamard NMR QIP, k experiments are performed, where k is the number of transitions of the observer qubit. In each of the 
k-1 experiments, the MF-$\pi$ pulse is applied on k/2 transitions (explained in text and fig.\ref{hada4}). 
The results of the k experiments can be suitably decoded, to obtain the output state of the computation, individually for each of the input states.
A two-dimensional spectrum is obtained by inserting the 
decoded data at suitable frequencies in the $F_1$ dimension, following the Fourier transform in the $F_2$ dimension.

(2) (a) Schematic energy level diagram of a three qubit system, and deviation populations of the equilibrium state. (b) The equilibrium spectrum of the 
observer qubit, whose transitions are labeled as the quantum states of other two qubits.

\begin{center}
\begin{figure}
\epsfig{file=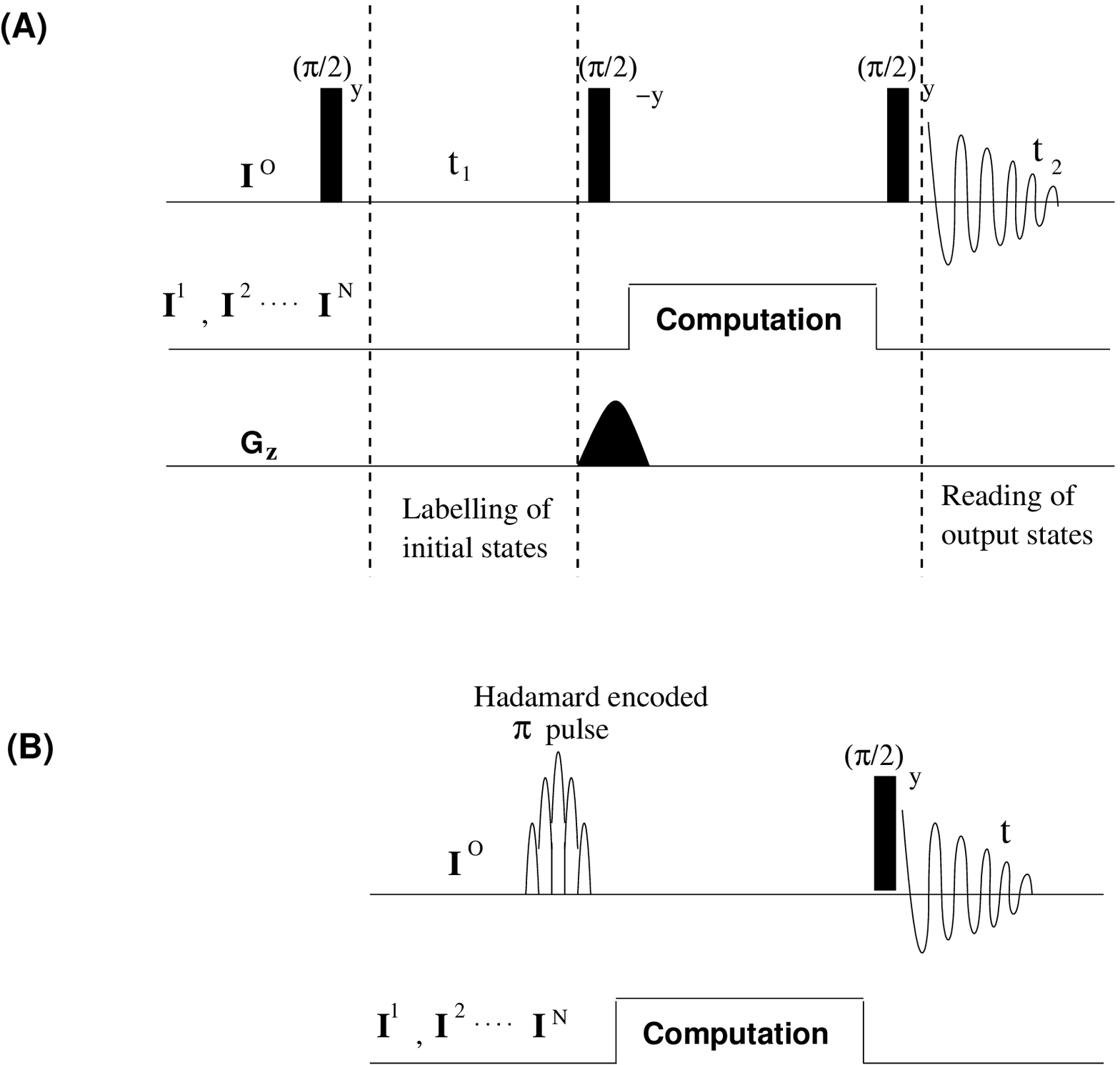,width=15cm}
\caption{} \label{pulse sequence}
\end{figure}
\end{center}

\begin{center}
\begin{figure}
\epsfig{file=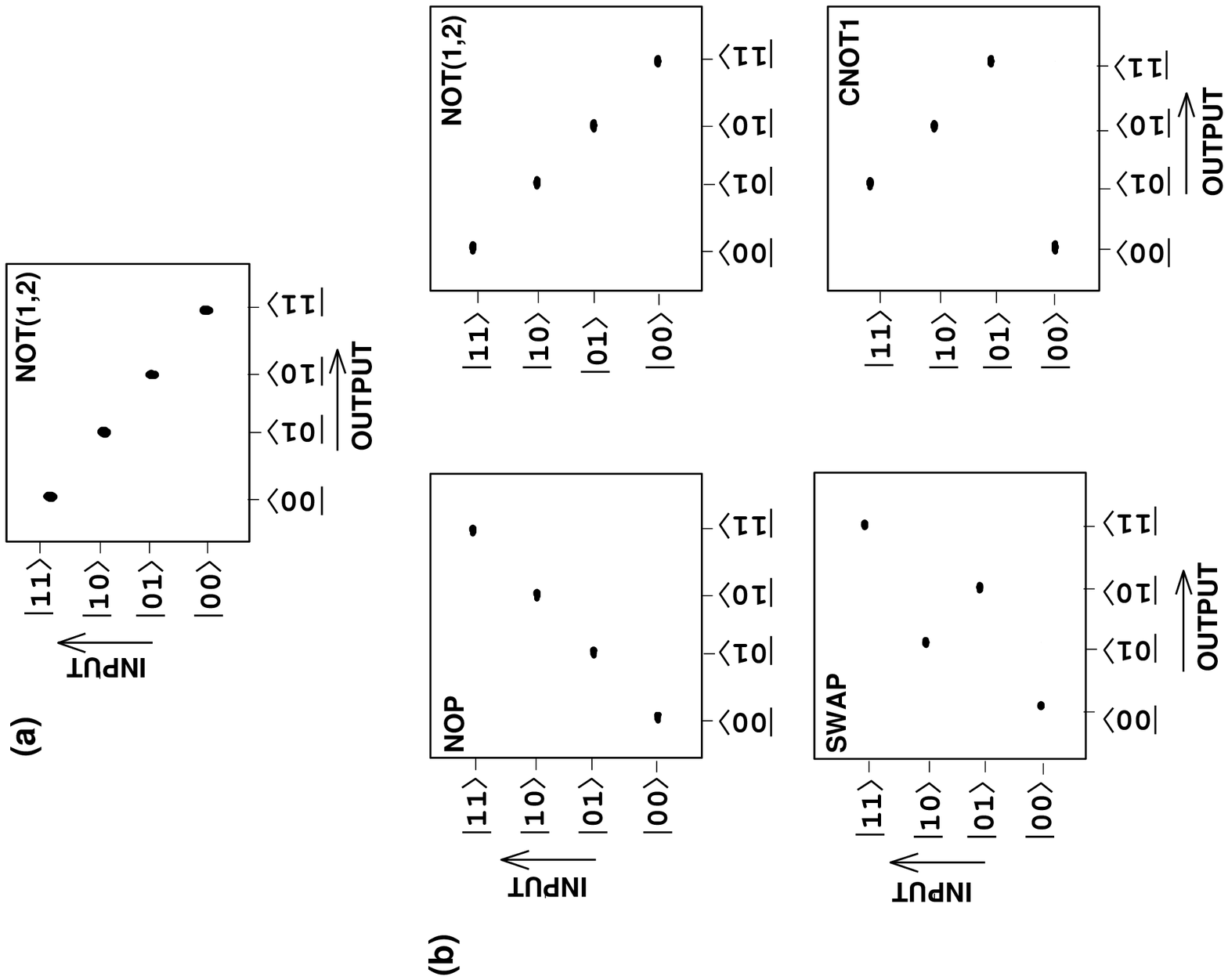,width=14cm,angle=270}
\caption{}   \label{2D of 3qubits}
\end{figure}
\end{center}

\begin{center}
\begin{figure}
\epsfig{file=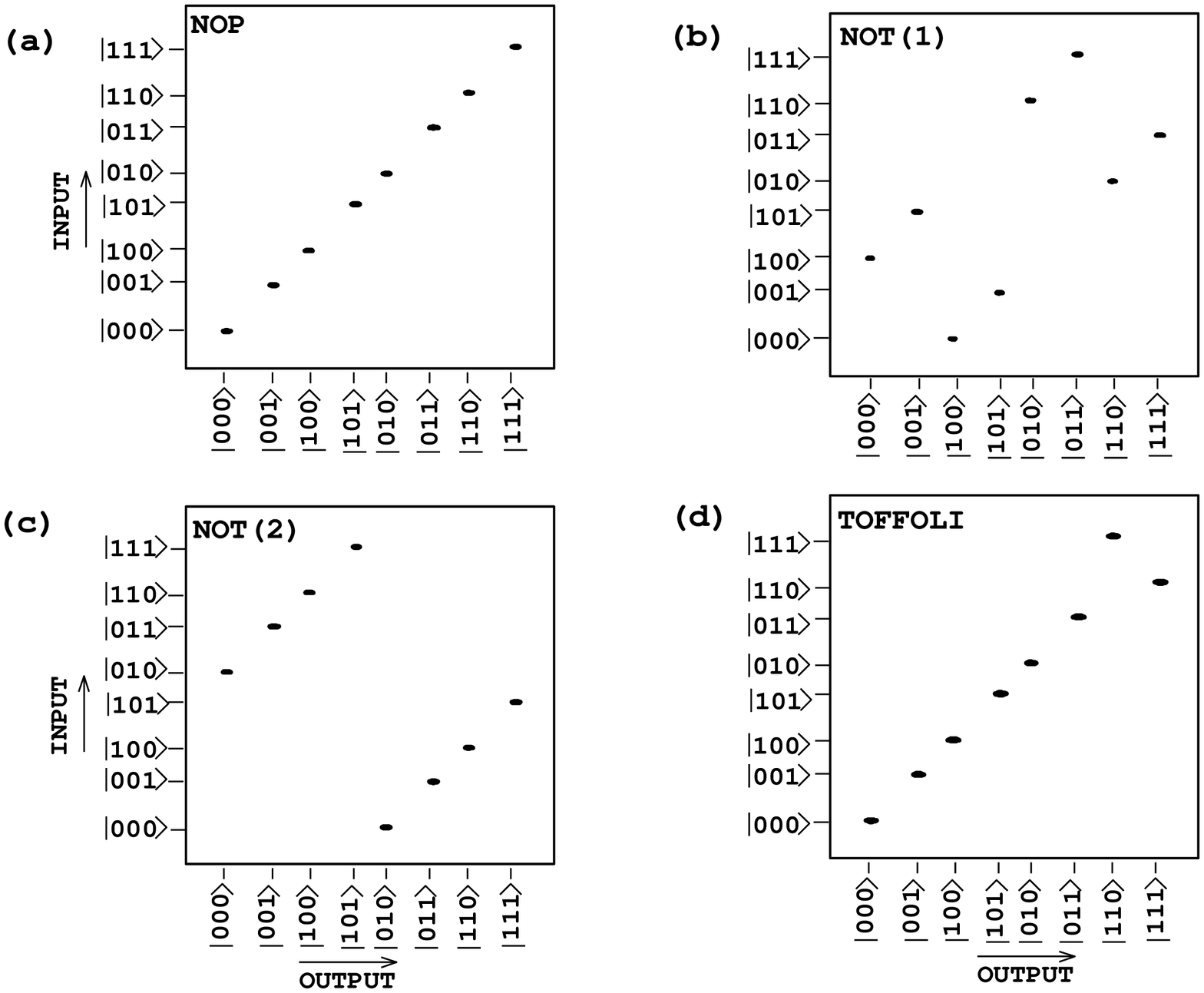,width=12cm,angle=270}
\caption{}    \label{2D of 4qubits}
\end{figure}
\end{center}

(3) (a) and (b) are Hadamard matrices, which are used to implement two and three qubit gates respectively 
(fig. \ref{2D of 3qubits}b, \ref{2D of 4qubits}), by using 2D Hadamard QIP (fig.\ref{pulse sequence}b). 
Each of the columns of the Hadamard matrix 
are assigned to the transitions of the observer qubit. 
In the matrix '+' and '-' corresponds to no pulse and $\pi$ pulse respectively. 
The product operators, associated with each of the encodings, are also given in the last column.

(4) Fluorine spectra of $C_2F_3I$. The three fluorines form a three qubit system. The chemical shifts of work qubits with 
respect to observer qubit, are 
$\Omega_1$=11807 Hz, $\Omega_2$= -17114 Hz, and the J-couplings are $J_{O1}$= 68.1 Hz, $J_{O2}$= -128.8 Hz, and $J_{12}$= 48.9 Hz. The transitions of the observer qubit $I^O$ represent the quantum states of the work qubits ($I^1$ and $I^2$). 
The relative signs were determined by selective spin tickling experiments \cite{ern}.

\begin{center}
\begin{figure}
\epsfig{file=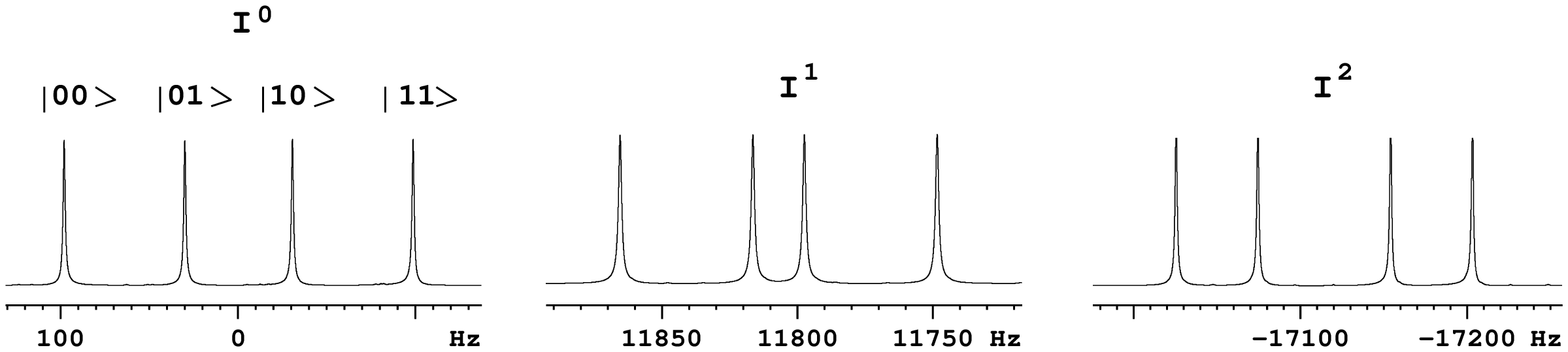,width=3cm,angle=270}
\caption{}   \label{1D of 3qubits}
\end{figure}
\end{center}

(5) (a) Implementation of two qubit NOT(1,2) gate, on a three qubit system (fig.\ref{1D of 3qubits}),by using 
conventional method given in fig.\ref{pulse sequence}A; 
128 $t_1$ increments are used, with 2 scans for each increment and a recycle delay of 20 sec., resulting in a total experimental 
time of 126 minutes. 
(b) Implementation of two qubit quantum gates by using 2D Hadamard QIP (fig.\ref{pulse sequence}b). Each 2D gate is recorded in four experiments, taking the total experimental time of less than 2 minutes. 
The encoding of the 
M.F. $\pi$ pulses, in each of the four experiments, is given in fig.\ref{hada4}a. NOP gate requires 
no pulse during the computation. NOT(1,2) is implemented by applying selective $\pi$ pulses on both the work qubits $I^1$ and $I^2$. Swap gate requires 
six transition selective $\pi$ pulses on transitions $\vert110\rangle$-$\vert111\rangle$, 
$\vert010\rangle$-$\vert011\rangle$, $\vert101\rangle$-$\vert111\rangle$, $\vert001\rangle$-$\vert011\rangle$, 
$\vert110\rangle$-$\vert111\rangle$ and $\vert010\rangle$-$\vert011\rangle$, CNOT(1) requires two selective $\pi$ pulses on transitions 
$\vert001\rangle$-$\vert011\rangle$ and $\vert101\rangle$-$\vert111\rangle$. The phases of the $\pi$ pulses during the computation, are set as (x, -x, -y, -y), in order to 
reduce the distortions due to pulse imperfections.

(6). Fluorine spectrum of tetra-fluro benzene. The four fluorines form a four qubit system, where $I^O$ is the observer qubit, whose transitions are 
labeled as the quantum states of the three work qubits ($I^1$, $I^2$ and $I^3$). The chemical shifts of work qubits with 
respect to observer qubit, are $\Omega_1$=13564.2 Hz, $\Omega_2$=6845.8 Hz, $\Omega_3$ = -5261.2 Hz, and the 
J-couplings are $J_{O1}$=10.5 HZ, $J_{O2}$=20.5 HZ, $J_{O3}$=6 HZ, 
$J_{12}$=9.5 HZ, $J_{13}$=22.7 HZ and $J_{23}$=21.9 HZ.

\begin{center}
\begin{figure}
\epsfig{file=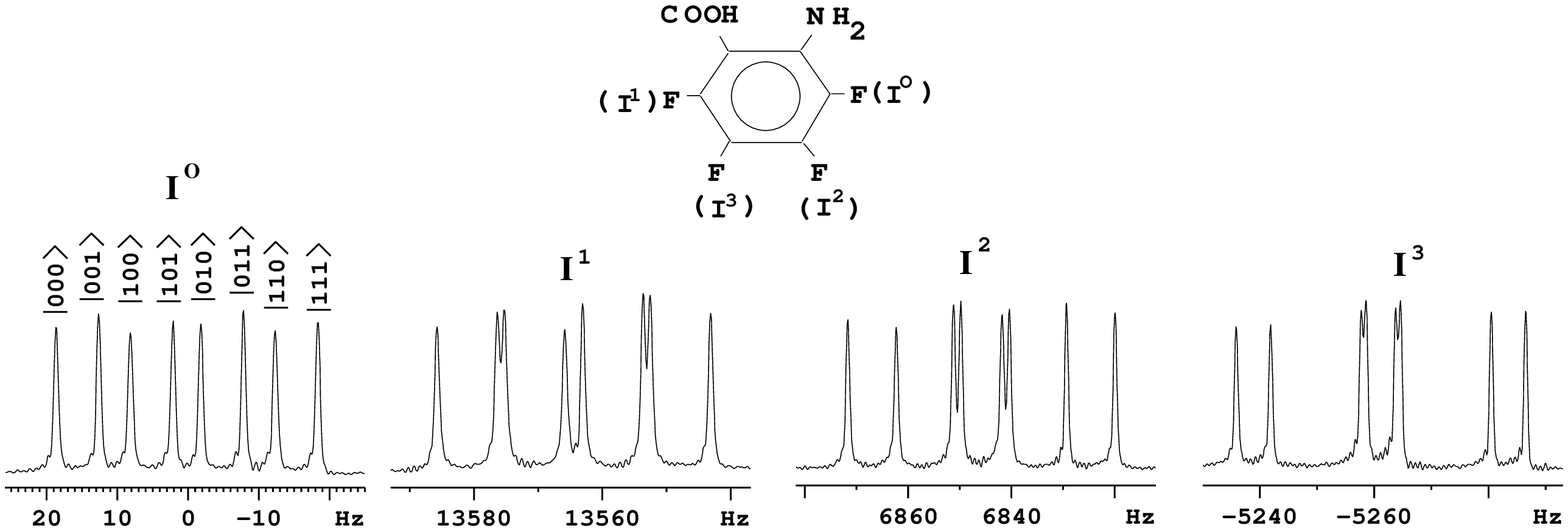,width=6cm,angle=270}
\caption{}   \label{1D of 4qubits}
\end{figure}
\end{center}

(7) Implementation of three qubit gates by using 2D Hadamard QIP (fig.\ref{pulse sequence}b), the 1D spectra of the observer qubit $I^O$ and the work 
qubits $I^1$, $I^2$ and 
$I^3$, are shown in fig.\ref{1D of 4qubits}. 
Each 2D spectrum is recorded in 8 experiments. The Hadamard encoding, in each of the 8 experiments, is achieved by 
J-evolution method (explained in text).
NOP gate is a unit matrix, hence the output states are same as 
the input states. NOT(1) and NOT(2), interchanges the states $\vert 0 \rangle$ and $\vert 1 \rangle$, of the 1'st and 2'nd qubits respectively. 
 Toffoli gate interchanges the states $\vert 0 \rangle$ and $\vert 1 \rangle$ of the third qubit ($I^3$), provided the other two work qubits ($I^1$ and 
$I^2)$ are in state $\vert 1 \rangle$.

\begin{center}
\begin{figure}
\epsfig{file=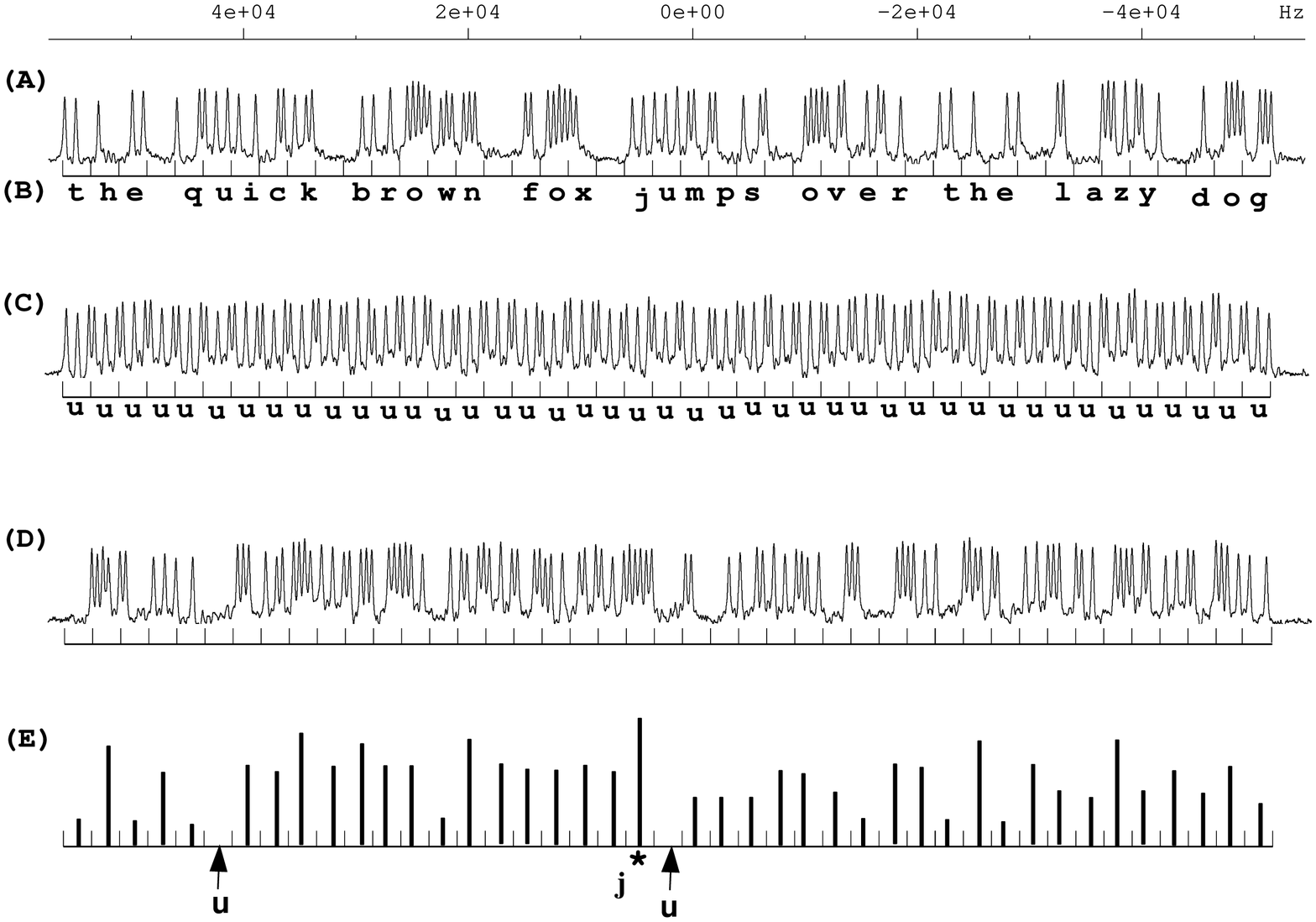,width=11cm,angle=270}
\caption{}   \label{fox}
\end{figure}
\end{center}

\begin{center}
\begin{figure}
\epsfig{file=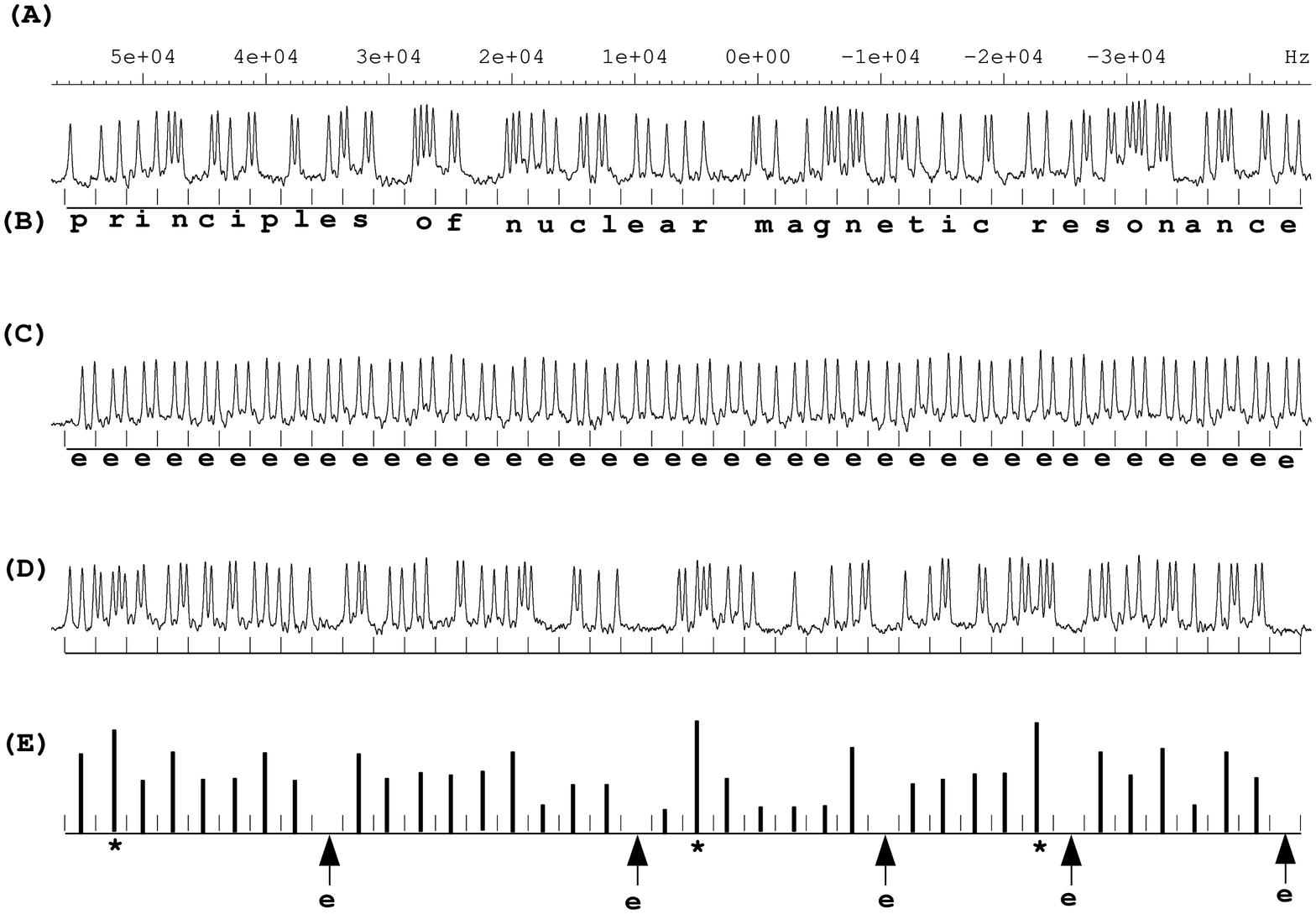,width=11cm,angle=270}
\caption{}   \label{pnmr}
\end{figure}
\end{center}

(8). Pulse sequence for the implementation of parallel search algorithm. The multi-frequency $\pi/2$ pulse is obtained by modulating the 
Gaussian pulse with 256 harmonics and the phase modulation for each of the harmonics is y or -y. 256 multi-frequency $\pi/2$ pulses are synthesized, 
which differ from each other, only in the phase modulation, which is according to the rows of 256-dimension Hadamard matrix, where + and - in the matrix 
corresponds to the phases y and -y respectively. 
The duration of the M.F. pulse is 30 m.s., and the gradient strength is 25 Gauss/cm. The 256 1D spectra obtained individually from 256 pulses, are independently stored for suitable decoding, for desired information and search as 
shown in Fig. \ref{fox} and \ref{pnmr}.

\begin{center}
\begin{figure}
\epsfig{file=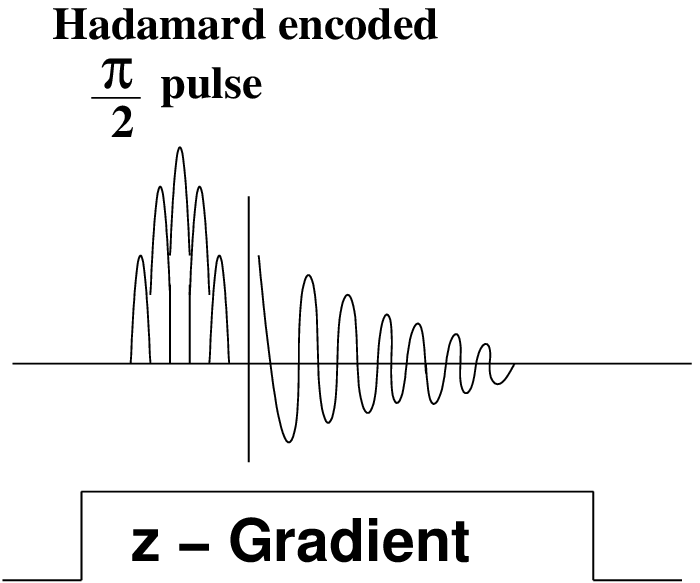,width=8cm}
\caption{}   \label{pulse sequence for spatial encoding}
\end{figure}
\end{center}

(9). (A) spectrum obtained from the Hadamard decoding of the 256 experiments (fig.\ref{pulse sequence for spatial encoding}), which represents the 
sentence (B), the quick brown fox jumps over 
the lazy dog. The sentence (B), consists of 215 bits or 43 ciphers, where each cipher is a five bit 
string, with a = 00001, b = 00010,....,z = 11100, and space = 00000. The "0" and "1" corresponds to "no excitation" and "excitation" respectively.  
(C) Ancilla pattern for letter u = 10101, "uuu....u (215 times). (D) The difference spectrum of (A) and (C). (E) The heights of the bars represent the 
integration of peak intensities of each of 
the five bit string of spectrum (D), the zero intensity (represented by arrow), indicates the presence of letter "u", and the intensity of letter "j" having a complimentary code 01010 (represented by *), is 4.92 units (theoretically 5 units). This method is known as XOR search.

(10). The Hadamard encoded data is also used to write another sentence, "principles of nuclear magnetic resonance"(B). An 
XOR search is performed for letter e = 00101 (C), and the results are given in (D) and (E). The letter z having the 
complimentary code 11010 is absent in (B). Hence there is no line of intensity 5 units in (E). The next intensity is of 4 units, 
which correspond to letters "j", "r" and "x", whose code differ by 4 units from that of letter "e". However, only letter "r" is present in "B", occurring thrice and marked by * in (E).

\newpage

\newpage

\newpage

\newpage

\newpage

\newpage

\newpage

\newpage

\newpage

\newpage

\end{document}